\documentclass[12pt,a4paper]{article}
\usepackage{amssymb,amsfonts,amsthm,amsmath}
\usepackage[mathscr]{eucal}

\usepackage{epsf}
\usepackage{graphicx}
\usepackage{epstopdf}
\usepackage{placeins}

\def\hang{\hangindent\parindent}
 \def\rf{\par\noindent\hang}

\DeclareMathAlphabet{\mathpzc}{OT1}{pzc}{m}{it}

\newcommand{\bD}{\boldsymbol{D}}
\newcommand{\ba}{\boldsymbol{a}}
\newcommand{\bH}{\boldsymbol{H}}
\newcommand{\bbeta}{\boldsymbol{\beta}}
\newcommand{\bI}{\boldsymbol{I}}
\newcommand{\bX}{\boldsymbol{X}}

\newcommand{\by}{\boldsymbol{y}}
\newcommand{\bzero}{\boldsymbol{0}}
\newcommand{\bc}{\boldsymbol{c}}

\newcommand{\pkstyle}[1]{\mbox{\footnotesize $#1$}}

\begin{document}

\baselineskip=20pt


\phantom{1}
\vspace{-1.5cm}

 \noindent {\Large {\bf  Upper bounds on the minimum coverage probability of
model averaged tail area confidence intervals
 in regression
}}


%
%
\begin{center}
\large{P{\normalsize \rm AUL} K{\normalsize \rm ABAILA} }
\end{center}
\begin{center}
{\large{\sl Department of Mathematics and Statistics}}
\end{center}
\begin{center}
{\large{\sl La Trobe University}}
\end{center}
%


\bigskip

\noindent \textit{Key words and phrases:} Model averaged confidence intervals; MATA confidence interval; minimum coverage probability.

\bigskip

\noindent \textit{MSC 2010:} Primary  62F25; secondary 62P12

\bigskip

\noindent \textit{Abstract:}  Frequentist model averaging has been proposed
as a method for incorporating ``model uncertainty'' into confidence interval construction. 
Such proposals have been of particular interest in the environmental and ecological statistics
communities.
A promising method of this type is the model averaged tail area (MATA) confidence interval put forward by Turek \& Fletcher, 2012.  
The performance of this interval depends greatly on the data-based model weights on which it is based.
A computationally convenient formula for the coverage probability of this interval is provided by 
Kabaila, Welsh and Abeysekera, 2016, in the simple scenario of two nested linear regression models.
We consider the more complicated scenario that there are many (32,768 in the example considered) linear regression models obtained as follows. 
 For each of a specified set of components of the regression parameter vector,
we either set the component to zero or let it vary
freely. 
We provide an easily-computed upper bound 
on the minimum coverage probability of the MATA confidence interval.
This upper bound provides evidence against the use of a model weight based on the Bayesian Information Criterion (BIC).

\bigskip

%
%
%
%
%
%
%
%
%
%


\noindent  {\bf 1. INTRODUCTION}

\medskip


\noindent Commonly in applied statistics, there is some uncertainty as to which explanatory variables should be included in 
the model. Frequentist model averaging has been proposed as a method for properly incorporating this ``model uncertainty'' into confidence interval construction. 
Such proposals have been of particular interest in the environmental and ecological statistics
communities, see e.g. Fieberg \& Johnson (2015, p.712) for a recent review.

The earliest approach to the construction of frequentist model averaged confidence intervals was to first construct a model averaged estimator of the parameter of 
interest as follows. This estimator is 
a data-based weighted average of the estimators of this parameter under the various models considered. In this approach, the model averaged confidence interval, with nominal coverage $1-\alpha$,  is centered on this
estimator and has width equal to the $1 - \alpha/2$ quantile of the standard normal distribution multiplied by an estimate of the standard deviation of this estimator
(Buckland {\it et al.}, 1997). However, Hjort \& Claeskens (2003, Section 4.3) show that the distributional assumption
 on which this confidence interval is based is completely incorrect in large samples.
 This problem effectively rules out the use of this confidence interval. 
 Hjort \& Claeskens (2003, equation 4.8) then propose a new frequentist model averaged confidence interval that has the desired minimum coverage probability in large samples. However, this interval is essentially the same as the standard confidence interval based on the full model (Kabaila \& Leeb, 2006, Remark 5b and Wang \& Zou, 2013).

 An important conceptual advance was made by 
 Fletcher \& Turek (2011) and Turek \& Fletcher (2012) who put forward the idea of 
 using data-based weighted averages across the models considered of {\sl procedures} for constructing confidence intervals. In this way the model averaged confidence interval is constructed in a single step, rather than first constructing a model averaged estimator, which is used as the center of this interval, and then seeking an appropriate formula for the width of this interval.
 However, some problems have been identified by Kabaila, Welsh \& Abeysekera (2016) with the 
 method of Fletcher \& Turek (2011). 
This leaves the model averaged tail area (MATA) confidence interval of Turek \& Fletcher (2012) as a promising method, particularly in the normal linear regression context since exactly pivotal quantities for the parameter of interest can be specified for each model under consideration. As Turek \& Fletcher (2102) note, their method can also be applied when one has only approximately pivotal quantities  for the parameter of interest for each model under consideration. However, the use of such approximately pivotal quantities (which may be obtained by via the parametric bootstrap) is outside the scope of the present paper.

Turek \& Fletcher (2012) considered a data-based weight on a model that is proportional to 
$\exp(-\text{AIC}/2)$, $\exp(-\text{AIC}_c/2)$ and $\exp(-\text{BIC}/2)$,
where AIC, $\text{AIC}_c$ and BIC are the Akaike Information Criterion, the Akaike Information Criterion
corrected for small samples and the Bayesian Information Criterion, respectively, for the model.
The performance of the MATA confidence interval depends greatly on the model weights on which it is based.
It is helpful to applied statisticians who wish to use MATA intervals if we can narrow down the choice of 
data-based model weight by eliminating the worst performing model weights from further consideration.

A computationally convenient formula for the exact coverage probability of the MATA interval is provided by 
Kabaila, Welsh \& Abeysekera (2016) in the simple scenario of two nested normal linear regression models: the full model and a submodel specified by a linear constraint on the regression parameter vector. They consider a parameter of 
interest that is a specified linear combination of the components of the regression parameter vector for the full model.  
Kabaila, Welsh \& Mainzer (2106) consider the same simple scenario 
in their evaluation of a MATA interval constructed using data-based weights based on Mallows' $C_P$.
Of course, it is of interest to also evaluate the MATA interval
in the more complicated situations that we average over more
than two ($2^{15}$ for the real life data considered in Section 5) normal linear regression models.

In the present paper, the family of models that we average over is obtained as follows. 
For each of a specified set of components of the regression parameter vector,
we either set the component to zero or let it vary
freely. For the MATA 
interval, we consider quite general data-based weights on these models.
These general weights include, as special cases, the weights considered by Turek \& Fletcher (2012)
and the weights based on Mallows' $C_P$ that are considered by Kabaila, Welsh \& Mainzer (2016).
Using the two new theorems presented in Section 3 of the present paper, we show how the results of Kabaila, Welsh \& Abeysekera (2016) 
can be used to provide a new easily-computed upper bound 
on the minimum coverage probability of the MATA 
interval
 in this situation.
This upper bound is analogous to the upper bounds of Kabaila \& Leeb (2006) and Kabaila \& Giri (2009) on the 
minimum coverage probability of the post-model-selection confidence interval in the context of the 
same family of models and is proved using the approach of Kabaila \& Giri (2009).

The most important measure (in the form of a single number) of the performance
of a confidence interval is its
{\sl confidence coefficient}, defined to be the
infimum of the coverage probability of a confidence interval (see e.g. Casella \& Berger, 2002, pp.418--419). If the {\sl confidence coefficient} of a confidence interval is far below its nominal coverage then this confidence interval should not be used. 
The main application of our new upper bound on the minimum coverage probability of the MATA interval is that it can be used to help eliminate poorly performing model weights
from further consideration.

Consider the linear regression model
 \begin{equation*}
 \by = \boldsymbol{X} \boldsymbol{\beta} + \boldsymbol{\varepsilon},
 \end{equation*}
where $\by$ is a random $n$-vector of responses, $\boldsymbol{X}$ is a known
$n\times p$ matrix with  linearly independent columns, $\boldsymbol{\beta}$ is
an unknown parameter $p$-vector and $\boldsymbol{\varepsilon} \sim \text{N}(0, \sigma^2 \bI)$
where $\sigma^2$ is an unknown positive parameter and $n > p$.
Suppose that
the quantity of interest is $\theta = \boldsymbol{a}^{\top} \boldsymbol{\beta}$
where $\boldsymbol{a}$ is a
specified non-zero $p$-vector. Our aim is to find a confidence
interval for $\theta$ with minimum coverage probability a pre-specified value
$1-\alpha$, based on an observation of $\by$.

Henceforth, let $\mathscr K$ denote the family of all subsets of $\{ q+1, \dots, p \}$ 
including the empty set, where $q$ is a specified integer satisfying $1 \le q <p$. 
For each $K \in \mathscr K$, let ${\cal M}_K$
denote the model for which $\beta_i = 0$ for all $i \in K$.
In other words, the number of models under consideration is $2^{p - q}$.
Suppose that the last $p - q$ components of $\boldsymbol{a}$ are zeros.  In other words, suppose that these models differ from each other 
only with respect to nuisance parameters, so that the quantity of interest $\theta$ has the same
meaning for all of these models.
This condition will commonly be satisfied, possibly after some minor reparametrization (see Section 5 for an example).
We consider quite general data-based weights on the models ${\cal M}_K$, where $K$ belongs to the 
family ${\mathscr K}$.
We then consider the MATA 
interval, with nominal coverage $1-\alpha$, obtained
by averaging over the these models
using these data-based weights. 
We denote this confidence interval by $I({\mathscr K})$.

Our easily-computed (calculated by repeated numerical evaluation of a double integral)
upper bound on the minimum coverage probability of
the MATA 
interval $I({\mathscr K})$ is obtained as follows.
We first prove the intuitively plausible result Theorem 2 (stated in Section 2) 
that the wider the class of models over which 
one averages using specified data-based model weights,
the smaller is the minimum coverage probability of the MATA 
interval, with nominal coverage $1-\alpha$.
Let $\widehat{\theta}$, $\widehat{\beta}_{q+1}, \ldots, \widehat{\beta}_p$ denote the least squares estimators
of $\theta$, $\beta_{q+1}, \ldots, \beta_p$ respectively.
Also let corr$\big(\widehat{\theta}, \widehat{\beta}_j \big)$ denote the correlation between $\widehat{\theta}$ and $\widehat{\beta}_j$, which is a known quantity that is determined by the design matrix $\bX$
and the vector $\ba$ which specifies the parameter of interest $\theta$.
It follows from the results of Kabaila, Welsh \& Abeysekera (2016) that the MATA 
interval, with nominal coverage
$1- \alpha$, obtained by data-based averaging over {\sl only} the full model and the submodel for which
$\beta_j = 0$ has minimum coverage probability that is the same decreasing function of  
$|\text{corr}\big(\widehat{\theta}, \widehat{\beta}_j \big)|$, for each $j \in \{ q+1, \dots, p \}$.
It follows from Theorem 1 that this minimum coverage probability is an upper bound on
the minimum coverage probability of the MATA 
interval $I({\mathscr K})$, 
for each $j \in \{ q+1, \dots, p \}$. 
Our upper bound on the minimum coverage probability of the MATA 
interval $I({\mathscr K})$ is 
simply the minimum of these upper bounds, which is attained for the value of $j \in \{ q+1, \dots, p \}$ maximizing 
$|\text{corr}\big(\widehat{\theta}, \widehat{\beta}_j \big)|$.
This upper bound depends on the design matrix $\bX$ and the vector $\boldsymbol{a}$ 
only through the known parameter 
$|\rho|_{\pkstyle{\rm max}}$
which we define to be the maximum over $j \in \{q +1, \dots, p\}$ of
$\big|\text{corr}\big(\widehat{\theta}, \widehat{\beta}_j \big)\big|$.
Since $|\rho|_{\pkstyle{\rm max}}$ is obtained by this maximization,
 it may be quite close to 1 in many applications.
We have written an {\tt R} computer program to evaluate this upper bound.

We use this computer program to provide evidence against the use of a 
data-based weight on the model ${\cal M}_K$ that is proportional to $\exp(-\text{BIC}(K)/2)$, 
where $\text{BIC}(K)$ denotes the BIC criterion for this model. Since AIC and BIC are similar 
criteria for $\ln(n)$ approximately equal to 2, we consider $n \ge 15$.
Figure 1 presents graphs of the upper bound on the minimum coverage probability of the MATA interval 
$I({\mathscr K})$,
with nominal
coverage 0.95, as a function of $|\rho|_{\pkstyle{\rm max}}$ for 
$p = 10$ and
$n \in \{15, 30, 70, 200 \}$. 
For each value of $n$ considered, this upper bound is found
to be a continuous decreasing function of $|\rho|_{\pkstyle{\rm max}}$
that falls well below $1-\alpha$ when $|\rho|_{\pkstyle{\rm max}}$ is close
to 1. Also, for each value of $|\rho|_{\pkstyle{\rm max}} > 0$
considered, this upper bound is found to be a decreasing function of $n$.
Figures similar to Figure 1 are presented in the Supplementary Material
for a wide range of values of $n$ and $p$. 
Figure 1 suggests the following large sample result: under the very weak condition that $|\rho|_{\pkstyle{\rm max}}$
converges to a positive number as $n \rightarrow \infty$, the minimum
coverage probability (i.e the {\sl confidence coefficient}) of the MATA interval $I({\mathscr K})$, with weight 
on model ${\cal M}_K$ proportional to $\exp(-\text{BIC}(K)/2)$, 
converges to 0 as $n \rightarrow \infty$. This suggested result turns out to be correct and is stated in Section 6.

Large sample results can have subtleties in their interpretation. These subtleties are briefly explored at the start of Section 6, before we state the main results of
this section. 
Our conclusion from these results and the Supplementary Material is that the MATA 
interval
with weight on the model ${\cal M}_K$ proportional to $\exp(-\text{BIC}(K)/2)$ should not be used
if $|\rho|_{\pkstyle{\rm max}}$ is not too far from 1 and $p/n$ is reasonably small, as judged from a figure, such as Figure 1, which is easily computed for any given $p$.



\FloatBarrier

%
\begin{figure}[h]
    \centering
    \includegraphics[scale=0.5]{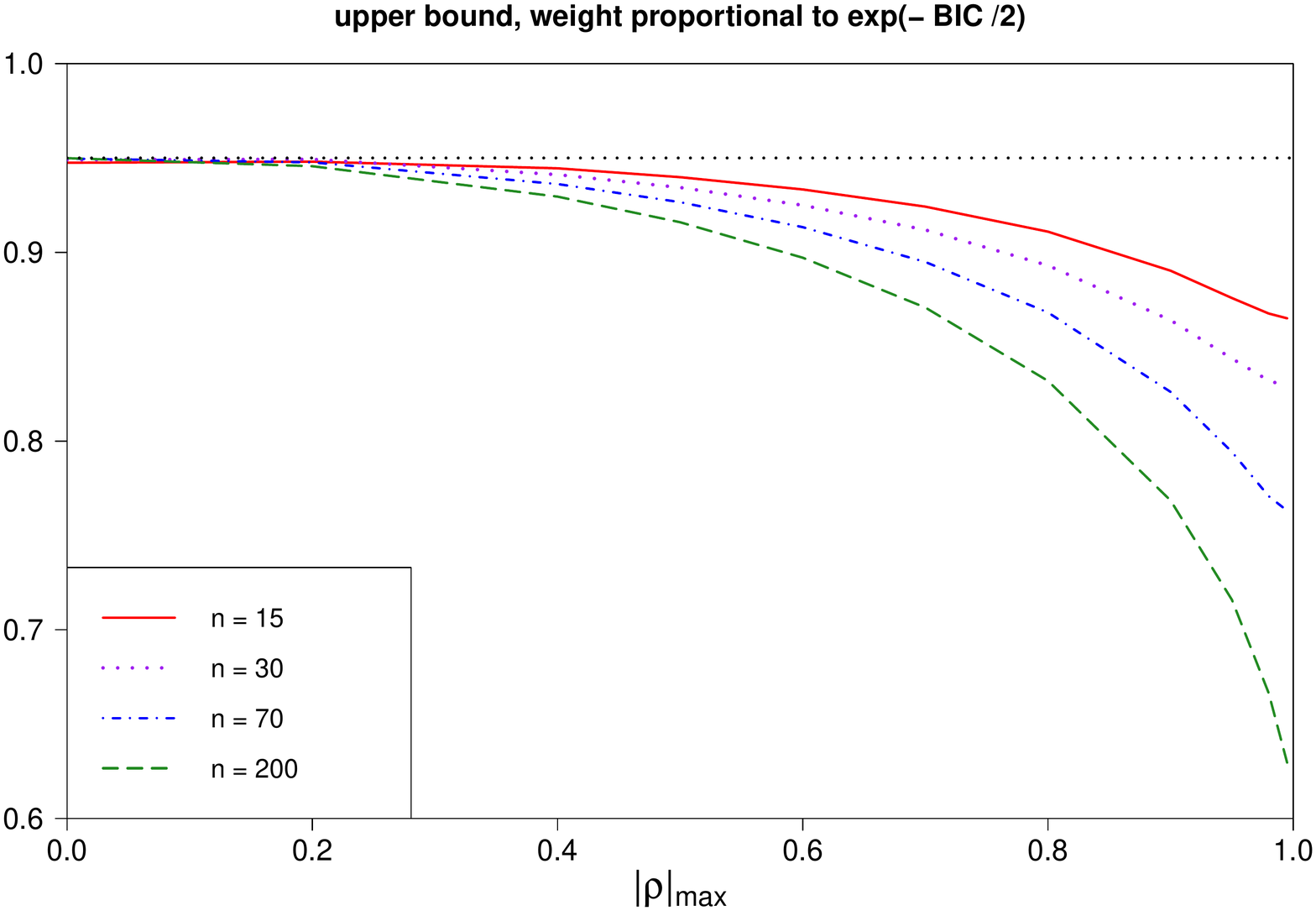}
   \caption{Graphs of the upper bound, described in Section 4,
    on the minimum coverage probability of the MATA 
confidence interval $I({\mathscr K})$, with nominal coverage 0.95, 
against $|\rho|_{\pkstyle{\rm max}}$. 
The weight on model ${\cal M}_K$  is proportional to
$\exp(-\text{BIC}(K)/2)$.
    Here  $p=10$ and $n = 15, 30, 70$ and 200. 
This plot includes a horizontal straight line with vertical
    axis intercept 0.95.}
\end{figure}

\FloatBarrier

\bigskip

\noindent {\bf 2. THE MATA 
INTERVAL FOR GENERAL DATA-BASED WEIGHTS}

\medskip

\noindent Let $\widehat{\boldsymbol{\beta}}$ denote the least-squares estimator of $\boldsymbol{\beta}$.
Let RSS denote the following residual sum of squares,
$$\text{RSS}=(\by - \boldsymbol{X} \widehat{\boldsymbol{\beta}})^{\top}
(\by - \boldsymbol{X} \widehat{\boldsymbol{\beta}}).$$
For each $K \in \mathscr K$,
let $\vert K\vert$ denote the
number of elements in $K$. Also, for
$K \ne \varnothing$,
let $\boldsymbol{H}_K$ denote the $\vert K
\vert \times p$ matrix whose $i$'th row consists of zeros except
for the $j$'th element which is 1, where $j$ is the $i$'th ordered
element of $K$. Thus $\boldsymbol{H}_K \boldsymbol{\beta} = \boldsymbol{0}$,
for the model ${\cal M}_K$ ($K \ne \varnothing$).
Let $\widehat{\boldsymbol{\beta}}_K$ denote
the least-squares estimator of $\boldsymbol{\beta}$ subject to this restriction.
Note that 
\begin{equation}
\label{FormulaBetaHatK}
\widehat{\boldsymbol{\beta}}_K
= \Big( \bI - (\bX^{\top} \bX)^{-1}   \bH_K^{\top}  \big(\bH_K (\bX^{\top} \bX)^{-1} \bH_K^{\top} \big)^{-1} \bH_K \Big)
\widehat{\bbeta}.
\end{equation}
Let $\text{RSS}_K$ denote the residual sum of squares
$$\text{RSS}_K =(\by - \boldsymbol{X} \widehat{\boldsymbol{\beta}}_K)^{\top}
(\by - \boldsymbol{X} \widehat{\boldsymbol{\beta}}_K)$$
and $S_K^2 = \text{RSS}_K/(n-p+|K|)$.
Also let $v(K) = \text{var}\big(\boldsymbol{a}^{\top} \widehat{\boldsymbol{\beta}}_K \big)/\sigma^2$,
where this variance is computed under the model ${\cal M}_K$.

We can choose a model from $\big \{ {\cal M}_K:  K \in {\mathscr K} \big\}$ by minimizing the following generalized information criterion
\begin{equation}
\label{GIC}
\text{GIC}(K) = n \ln (\text{RSS}_K) + d(p-\vert K\vert )
\end{equation}
with respect to $ K \in {\mathscr K}$,
where $d$ is a nonnegative number ($d =2$ for AIC and $d = \ln(n)$ for BIC)
and $\text{RRS}_K = \text{RSS}$ for $K = \varnothing$.
A weight for model ${\cal M}_K$ ($K \in {\mathscr K}$) that is proportional to
$\exp(-\text{GIC}(K)/2)$, for either $d = 2$ or $d = \ln(n)$, was considered
by Turek \& Fletcher (2012).

We introduce quite general forms of model weights based on the statistics $U_K / \text{RSS}$,
where
\begin{equation*}
U_K
= \big(\boldsymbol{H}_K \widehat{\boldsymbol{\beta}} \big)^{\top}
\big(\boldsymbol{H}_K (\boldsymbol{X}^{\top}\boldsymbol{X})^{-1} \boldsymbol{H}_K^{\top} \big)^{-1}
\boldsymbol{H}_K \widehat{\boldsymbol{\beta}},
\qquad K \in {\mathscr K} \setminus \{\varnothing \}.
\end{equation*}
Some motivation for the use of such weights is provided by the fact that
\begin{equation*}
\frac{U_K / |K|}{\text{RSS} / (n-p)}
\end{equation*}
is the usual test statistic for testing the null hypothesis that $\boldsymbol{H}_K \boldsymbol{\beta} = \boldsymbol{0}$
against the alternative hypothesis that $\boldsymbol{H}_K \boldsymbol{\beta} \ne \boldsymbol{0}$. This test statistic has
an $F_{|K|, n-p}$ distribution under this null hypothesis.
Obviously, $U_K / \text{RSS} = V_K \big/ (\text{RSS}/\sigma^2)$, where
\begin{equation*}
V_K
= \big(\boldsymbol{H}_K (\widehat{\boldsymbol{\beta}}/\sigma) \big)^{\top}
\big(\boldsymbol{H}_K (\boldsymbol{X}^{\top}\boldsymbol{X})^{-1} \boldsymbol{H}_K^{\top} \big)^{-1}
\boldsymbol{H}_K (\widehat{\boldsymbol{\beta}}/\sigma),
\qquad K \in {\mathscr K} \setminus \{\varnothing \}.
\end{equation*}
Now, for any given $K \in {\mathscr K} \setminus \{\varnothing \}$, $V_K$ and $\text{RSS}/\sigma^2$ are independent random variables,
where $\text{RSS}/\sigma^2 \sim \chi^2_{n-p}$ and $V_K$
has a noncentral chi-squared distribution with degrees of freedom $|K|$ and noncentrality parameter
\begin{equation}
\label{NoncentralityParameter}
\lambda = (1/2) \big(\boldsymbol{H}_K (\boldsymbol{\beta}/\sigma) \big)^{\top}
\big(\boldsymbol{H}_K (\boldsymbol{X}^{\top}\boldsymbol{X})^{-1} \boldsymbol{H}_K^{\top} \big)^{-1}
\boldsymbol{H}_K (\boldsymbol{\beta}/\sigma),
\end{equation}
see e.g. Graybill (1976, p.127).
Thus $U_K / \text{RSS}$ may be viewed as a data-based measure of the deviation of the model ${\cal M}_K$ from
the true model. This suggests a data-based weight $w(K;  {\mathscr K})$ on the model ${\cal M}_K$ 
($K \in {\mathscr K}$) 
given by
\begin{equation}
\label{weight}
\begin{split}
w(K;  {\mathscr K}) =
\begin{cases}
\displaystyle{\frac{1}{1 + \sum_{L \in {\mathscr K} \setminus \{\varnothing \}} r(U_L / \text{RSS}, |L|)}} &\text{for}\ \ \ K = \varnothing \\
\\
\displaystyle{\frac{r(U_K / \text{RSS}, |K|)}{1 + \sum_{L \in {\mathscr K} \setminus \{\varnothing \}} r(U_L / \text{RSS}, |L|)}}
&\text{otherwise}.
\end{cases}
\end{split}
\end{equation}
Here, the function $r: (0, \infty) \times \{1, \dots, p-q\} \rightarrow (0, \infty)$ satisfies the following conditions:

\begin{enumerate}


\item[\textbf{C1}] For each $y \in \{1, \dots, p-q\}$, $r(x,y)$ is a continuous decreasing function of
$x$ that approaches 0 as $x \rightarrow \infty$.

\item[\textbf{C2}]  For each $x \in (0, \infty)$, $r(x,y)$ is an increasing function of $y \in \{1, \dots, p-q\}$.

\end{enumerate}

\noindent The motivation for the second of these conditions is as follows.
According to \eqref{weight}, the weight on model ${\cal M}_K$ is proportional
to $r(U_K / \text{RSS}, |K|)$, where $U_K / \text{RSS}$ is a data-based measure
of the deviation of the model ${\cal M}_K$ from the true model and $|K|$ is 
the number of regression parameters that are set to 0. We want 
$r(U_K / \text{RSS}, |K|)$ to be an increasing function of $|K|$ since
this leads to $r(U_K / \text{RSS}, |K|)$ being a decreasing function of $p - |K|$,
which is the number of regression parameters in the model
${\cal M}_K$.
As shown in the appendix, a weight for model ${\cal M}_K$ ($K \in {\mathscr K}$) that is proportional to
$\exp(-\text{GIC}(K)/2)$ has the form described by \eqref{weight} above.


The MATA 
interval $I({\mathscr K})$ for $\theta$, with nominal coverage $1-\alpha$
 and obtained
by averaging (using the data-based weights \eqref{weight}) over the models $\big \{ {\cal M}_K:  K \in {\mathscr K} \big\}$
 is obtained as follows. Let
\begin{equation}
\label{hzyK}
h \big(z, \by; {\mathscr K} \big) 
= \sum_{K \in {\mathscr K}} w(K;  {\mathscr K}) \,
G_{n - p + |K|} \left( \frac{\boldsymbol{a}^{\top} \widehat{\boldsymbol{\beta}}_K - z}{S_K \, (v(K))^{1/2}}  \right),
\end{equation}
where $G_{\nu}$ is the $t_\nu$ cdf.
The MATA 
interval $I({\mathscr K}) = \big[ \widehat{\theta}_{\ell},  \widehat{\theta}_u  \big]$, is 
obtained by solving
\begin{equation}
\label{MATA_EstimatingEqns}
h \big(\widehat{\theta}_{\ell}, \by; {\mathscr K} \big) = 1 - \alpha/2 
\ \ \text{and} \ \ h \big(\widehat{\theta}_u, \by; {\mathscr K} \big) = \alpha/2
\end{equation}
for $\widehat{\theta}_{\ell}$ and $\widehat{\theta}_u$.

\bigskip

\noindent  {\bf 3. TWO IMPORTANT PRELIMINARY RESULTS}

\medskip

\noindent  Remember the following definitions given in the introduction. Let $\mathscr K$ denote the family of all subsets of $\{ q+1, \dots, p \}$ ($1 \le q < p$),
including the empty set.
For each $K \in \mathscr K$, let ${\cal M}_K$
denote the model for which $\beta_i = 0$ for all $i \in K$.
Let $I({\mathscr K})$ denote
the MATA 
interval, with nominal coverage $1-\alpha$,  obtained
by averaging (using the data-based weights \eqref{weight}) over the models $\big \{ {\cal M}_K:  K \in {\mathscr K} \big\}$.
Throughout this section we assume that $\ba$, $\bX$  and $q$ are given.
Remember, we assume that the last $p-q$ components of $\ba$ are zeros.
The following lemma, proved in the appendix, paves the way for Theorems 1 and 2, which are the main results of this section.

\medskip

\noindent {\bf Lemma 1.} \  {\sl For each given $K \in {\mathscr K} \setminus \{\varnothing\}$,
\begin{equation}
\label{SecondTermEstEqn}
G_{n - p + |K|} \left( \frac{\boldsymbol{a}^{\top} \widehat{\boldsymbol{\beta}}_K - \ba^{\top} \bbeta}{S_K \, (v(K))^{1/2}}  \right)
\end{equation}
can be expressed
as a function of $(\widehat{\bbeta} - \bbeta) / \sigma$, ${\rm RSS} / \sigma^2$ and the variables 
in the set $\{ \beta_i / \sigma: i \in K \}$. Also, for $K = \varnothing$, \eqref{SecondTermEstEqn}
can be expressed
as a function of $(\widehat{\bbeta} - \bbeta) / \sigma$ and ${\rm RSS} / \sigma^2$.
%
}

\medskip

It is intuitively plausible that the wider the class of models over which one averages using specified data-based model weights,
the smaller is the minimum coverage probability of the MATA 
interval, with nominal coverage $1-\alpha$.
Theorem 2 below formalizes this plausible result.
Suppose that the integer $\ell$ satisfies $q+1 < \ell < p$. 
Let ${\mathscr K}^{**}$ denote the family of all subsets of $\{ \ell+1, \dots, p \}$,
including the empty set. Obviously, 
${\mathscr K}^{**} \subset {\mathscr K}$.
Let $I({\mathscr K}^{**})$ denote
the MATA 
interval, with nominal coverage $1-\alpha$,  obtained
by averaging 
\big(using the data-based weights \eqref{weight}, but with ${\mathscr K}$ replaced by ${\mathscr K}^{**}$ \big) 
over the models 
$\big \{ {\cal M}_K:  K \in {\mathscr K}^{**} \big\}$.
The following theorem is a necessary preliminary to Theorem 2.

\medskip


\noindent {\bf Theorem 1.} 
\vspace{-0.3cm}
{\sl 

\begin{enumerate}

\item

The coverage probability of the MATA 
interval $I({\mathscr K})$,  \newline
$P_{\boldsymbol{\beta}, \sigma}(\theta \in I({\mathscr K}))$, 
is a function of $(1/\sigma)(\beta_{q+1},\ldots, \beta_p)$.

\item

The coverage probability of the MATA 
interval $I({\mathscr K}^{**})$, \newline
$P_{\boldsymbol{\beta}, \sigma}(\theta \in I({\mathscr K}^{**}))$,
is a function of $(1/\sigma)(\beta_{\ell+1},\ldots, \beta_p)$.

\end{enumerate}

}

\smallskip

\noindent The proofs of parts (a) and (b) of this theorem are virtually identical and so only part (a) is proved in 
the appendix.

We will use the following theorem (proved in the appendix) in Section 4 to describe an easily-computed upper bound on the minimum coverage
probability of  the MATA 
interval $I({\mathscr K})$.

\bigskip

\noindent {\bf Theorem 2.} {\sl The minimum coverage probability of the MATA 
interval $I({\mathscr K})$, 
with nominal coverage $1-\alpha$, obtained
by averaging (using the data-based weights \eqref{weight}) over the models 
$\big \{ {\cal M}_K:  K \in {\mathscr K} \big\}$
is bounded above by the minimum over $(1/\sigma)(\beta_{\ell+1},\ldots, \beta_p) \in {\mathbb R}^{p -\ell}$ of
\begin{equation*}
P \big (  \theta \in  I({\mathscr K}^{**}) \big),
\end{equation*}
where $I({\mathscr K}^{**})$ denotes  the MATA 
interval,
with nominal coverage $1-\alpha$, obtained
by averaging \big(using the data-based weights \eqref{weight}, but with ${\mathscr K}$ replaced by ${\mathscr K}^{**}$ \big) over the models 
$\big \{ {\cal M}_K:  K \in {\mathscr K}^{**} \big\}$.
}

%

%

\newpage



\noindent { \bf {4. AN EASILY-COMPUTED UPPER BOUND ON THE
MINIMUM COVERAGE PROBABILITY OF THE MATA 
INTERVAL}}

\medskip

\noindent In this section we present an easily-computed upper bound on the minimum coverage probability of the
MATA 
interval $I({\mathscr K})$, with nominal coverage $1-\alpha$, obtained
by averaging (using the data-based weights \eqref{weight}) over the models $\big \{ {\cal M}_K:  K \in {\mathscr K} \big\}$.
Assume, 
for notational convenience,
that $\big|\text{corr}\big(\widehat{\theta}, \widehat{\beta}_j \big)\big|$
is maximized with respect to
$j \in \{q +1, \dots, p\}$ at $j=p$.
This assumption can always be satisfied using, if necessary, an initial rearrangement
of the order of the last $p - q$ columns of the matrix $\bX$.
Theorem 2 implies that
this minimum coverage probability is bounded above
by the coverage probability of the
MATA 
interval $I({\mathscr K}^*)$, with nominal coverage $1-\alpha$, 
for 
${\mathscr K}^* = \big \{ \varnothing, \{ p \}   \big \}$
and any given  $\beta_p/\sigma$.
Theorem 1 of Kabaila, Welsh \& Abeysekera (2016) provides a computationally-convenient expression for the latter coverage probability.
This expression is easily minimized numerically with respect to  $\beta_p/\sigma$ to obtain the value of an 
upper bound on the minimum coverage probability of the
MATA 
interval $I({\mathscr K})$, with nominal coverage $1-\alpha$.

To apply Theorem 1 of Kabaila, Welsh \& Abeysekera (2016), we introduce the following notation. Let $\bc$ be the $p$-vector
$(0, \dots, 0, 1)$, whose first $p-1$ components are zeros. Also let  $\widehat{\sigma}^2 = \text{RSS}/(n-p)$, 
$v_{\theta} = \text{var}(\widehat{\theta})/\sigma^2 = \ba^{\top} (\bX^{\top} \bX)^{-1} \ba$,
$v_p = \text{var}(\widehat{\beta}_p)/\sigma^2 = \bc^{\top} (\bX^{\top} \bX)^{-1} \bc$
and $\gamma = \beta_p / (\sigma \, v_p^{1/2})$. 
Observe that $\gamma$ is a scaled version of $\beta_p$. This scaling is very
helpful for the computation of the minimum coverage probability of the MATA interval,
as this minimum coverage is achieved at roughly the same value of $\gamma$, for small and
moderate sample sizes $n$.
Define
$\widetilde{\rho} = \text{corr}(\widehat{\theta}, \widehat{\beta}_p)$, which is equal to
$\ba^{\top} (\bX^{\top} \bX)^{-1} \bc \big / (v_{\theta} \, v_p)^{1/2}$.
Note that 
$v_{\theta}$, $v_p$ and $\widetilde{\rho}$ are known, whereas $\gamma$ is an unknown parameter.
Also note that $|\widetilde{\rho}| = |\rho|_{\pkstyle{\rm max}}$.
Finally, let $m = n - p$.

It follows from \eqref{weight}
that the weight $w \big({\{ p \}}, {\mathscr K}^* \big)$ on the model ${\cal M}_{\{ p \}}$
is given by 
\begin{equation*}
w\big({\{ p \}}, {\mathscr K}^* \big)
= 1 \Bigg /
\left(1 + 
\displaystyle{\frac{1}{r \left(\widehat{\beta}_p^2/(m \widehat{\sigma}^2 v_p) , 1 \right)}}
\right).
\end{equation*}
Therefore, the function $w_1$ defined by Kabaila {\it et al.} (2016)
must satisfy
\begin{equation*}
w_1 \left( \frac{\widehat{\beta}_p^2}{\widehat{\sigma}^2 v_p} \right) 
= w\big({\{ p \}}, {\mathscr K}^* \big),
\end{equation*}
so that 
\begin{equation*}
w_1(z)
= 1 \bigg / 
\left(1 + 
\displaystyle{\frac{1}{r \left(z/m , 1 \right)}}
\right).
\end{equation*}
%
Condition \textbf{C1} on the function $r$ implies that $w_1: [0, \infty) \rightarrow [0,1]$ is a decreasing continuous
function, 
such that $w_1(z)$ approaches 0 as $z \rightarrow \infty$.
For the particular case that the weight on the model ${\cal M}_K$ is proportional to
$\exp(-\text{GIC}(K)/2)$, as shown in the appendix, $r(x, 1) = \exp(d /2) \big/(1 + x)^{n/2}$
and consequently
\begin{equation}
\label{Definition_w1}
w_1(z) 
= \frac{1}{1 + \left( 1 + \frac{z}{m}\right)^{n/2} \exp(-d/2)}.
\end{equation}

We now apply the results of Kabaila, Welsh \& Abeysekera (2016).
The function $\delta_u(x,y)$ is defined on page 4 of this paper. As shown on page 6 of this paper, for the scenario considered in the present paper, this function takes the following particular form.
For  $0 < u < 1$, define $\delta_u(x,y)$ to be the solution for $\delta$ in the equation
\begin{equation*}
w_1(x^2/y^2) \, G_{m+1} \left( \left( \frac{m +1}{x^2 + m y^2}\right)^{1/2} 
\frac{\delta - \widetilde{\rho} \, x}{(1 - \widetilde{\rho}^{\, 2})^{1/2}} \right)
+ \big(1 - w_1(x^2/y^2) \big) \, G_m(\delta /  y) = u,
\end{equation*}
where  $G_{\nu}$ denotes the $t_{\nu}$ cdf. An immediate consequence of Theorem 1 of Kabaila, Welsh \& Abeysekera (2016) is that
 the coverage probability of the
MATA 
interval $I({\mathscr K}^*)$, with nominal coverage $1-\alpha$, 
and any given  $\gamma$ is given by 
\begin{equation}
\label{DoubleIntegral}
\int_0^{\infty} \int_{- \infty}^{\infty} 
\left( \Phi \left( \frac{\delta_{1 - \alpha/2}(x,y) - \widetilde{\rho}(x - y)}{(1 - \widetilde{\rho}^{\, 2})^{1/2}}\right)  
- \Phi \left( \frac{\delta_{\alpha/2}(x,y) - \widetilde{\rho}(x - y)}{(1 - \widetilde{\rho}^{\, 2})^{1/2}}\right)  \right) \phi(x - \gamma) f_m(y) \, dx \, dy,
\end{equation}
where $\Phi$ and $\phi$ denote the $N(0,1)$ cdf and pdf, respectively, and $f_{\nu}$ denotes the pdf
of $(Q / \nu)^{1/2}$, where $Q \sim \chi^2_{\nu}$. 
As noted on page 6 of Kabaila, Welsh \& Abeysekera (2016), the conditions required 
for Theorem 3 of Kabaila, Welsh \& Abeysekera (2016) to hold are satisfied.
This theorem
implies that this coverage probability is an even function of $\gamma$ for fixed
$\widetilde{\rho}$ and an even function of $\widetilde{\rho}$ for fixed $\gamma$. The
upper bound on the minimum coverage probability of the
MATA 
interval $I({\mathscr K})$, with nominal coverage $1-\alpha$, is obtained by 
setting $\widetilde{\rho} = |\rho|_{\pkstyle{\rm max}}$ and then
minimizing 
\eqref{DoubleIntegral} over $\gamma \ge 0$.
The double integral \eqref{DoubleIntegral} is very easily computed using the methods
described in Appendix B of Kabaila, Welsh \& Mainzer (2016).
An {\tt R} computer program for the computation of this double integral is available upon request.


\bigskip

\noindent { \bf {5. NUMERICAL ILLUSTRATIONS}}

\medskip

\noindent In this section, we present some computed values of the upper bound, described in the previous section,
on the minimum coverage probability of the
MATA 
interval $I({\mathscr K})$, with nominal coverage 0.95, obtained
using a weight for model ${\cal M}_K$ ($K \in {\mathscr K}$) that is proportional to
$\exp(-\text{GIC}(K)/2)$ for both $d = 2$ (AIC) and $d = \ln(n)$ (BIC).
Consider the real life Air Pollution data described in 
Section 11.14 of Chatterjee \& Hadi (2012). The purpose of collecting this data was to study
the dependence of total mortality on climate, socioeconomic and pollution explanatory variables. 
Let $z_{i+1}$ denote the explanatory variable $X_i$ 
described in Table 11.11 of Chatterjee \& Hadi (2012), for $i = 1, \dots, 15$.
Consider the following linear regression model for this data:
\begin{equation*}
y = \psi + \beta_2 z_2 + \dots + \beta_{16} z_{16} + \varepsilon,
\end{equation*}
where the response variable $y$ is the total age-adjusted mortality from all causes, 
$\psi, \beta_2, \dots, \beta_{16}$ are unknown parameters and 
$\varepsilon \sim N(0, \sigma^2)$, for $\sigma^2$ an unknown parameter.
In this case, $n = 60$ and $p = 16$. 
Suppose that $\mathscr K$ is the family of all subsets of $\{2, \dots, 16 \}$ 
including the empty set. 
For each $K \in \mathscr K$, let ${\cal M}_K$
denote the model for which $\beta_i = 0$ for all $i \in K$.
In other words, the number of models under consideration is $2^{15} = 32,768$.
Suppose that the parameter of interest $\theta$ is $E(y)$ for 
$(z_2, \dots, z_{16}) = (z_2^*, \dots, z_{16}^*)$, where $(z_2^*, \dots, z_{16}^*)$
is equal to
\begin{equation*}
(37.37, 33.98, 74.58, 8.8, 3.26, 10.97, 80.91, 3876.05, 11.87, 46.08, 14.37, 100, 30, 140, 57.57).
\end{equation*}
Note that $(z_2^*, \dots, z_{16}^*)$ is well within the range of the values of $(z_2, \dots, z_{16})$
in the data. Obviously, $\theta = \psi + \beta_2 z_2^* + \dots + \beta_{16} z_{16}^*$ and so 
\begin{equation*}
y = \beta_1 + \beta_2 (z_2 - z_2^*)+ \dots + \beta_{16} (z_{16} - z_{16}^*) + \varepsilon
\end{equation*}
and $\theta = \beta_1$. In this parametrization of the linear regression model, $\theta$ has the 
same meaning for all the models ${\cal M}_K$, where $K \in \mathscr K$.
In this case, $|\rho|_{\pkstyle{\rm max}} = 0.9599$ and the upper bound on the minimum coverage probability of  
the MATA 
interval $I({\mathscr K})$, with nominal coverage 0.95,
is (a) 0.8900 for $d = 2$ (AIC) and (b) 0.7940 for $d = \ln(n)$ (BIC).


\bigskip

\noindent { \bf {6. LARGE SAMPLE RESULTS FOR THE MATA INTERVAL}}

\medskip

The main result of this section provides conditions under which the MATA interval $I({\mathscr K})$, with weight 
on model ${\cal M}_K$ proportional to $\exp(-\text{GIC}(K)/2)$, 
has minimum
coverage probability (i.e. {\sl confidence coefficient}) that converges to 0 as $n \rightarrow \infty$. 
An important advantage of the results presented in Sections 1--5 is that they are exact finite sample results and consequently their interpretation is very straightforward. By contrast, large sample results can have subtleties in their interpretation. It is these subtleties that we briefly explore before stating the main results of
this section. We begin by reminding the reader of Hodges's superefficient estimator and the well-known subtleties in the interpretation of large sample results for this point estimator. We then note that similar subtleties in the interpretation of large sample results also occur in the context of confidence intervals. Finally, we present the 
main result of this section which concerns the 
MATA interval.

Hodges's superefficient estimator is described as follows. Suppose that $X_1, X_2, \dots$ are independent and identically $N(\theta, 1)$ distributed, where
$\theta \in \Theta = \mathbb{R}$. The usual estimator of $\theta$ is  
$\overline{X}_n = (\sum_{i=1}^n X_i) / n$. 
Of course, $n E \big( (\overline{X}_n - \theta)^2 \big) = 1$ for all $\theta \in \Theta$.
Hodges's superefficient estimator is
\begin{equation*}
T_n =
\begin{cases}
\overline{X}_n &\text{if}\ \  |\overline{X}_n| > n^{-1/4}
\\
b \overline{X}_n &\text{if}\ \ |\overline{X}_n| \le n^{-1/4},
\end{cases}
\end{equation*}
where $0 < b < 1$. 
As shown on p.442 of Lehmann and Casella 
(1998), $\lim_{n \rightarrow \infty} n E \big( (T_n - \theta)^2 \big) = 1$ if $\theta \ne 0$ and 
$\lim_{n \rightarrow \infty} n E \big( (T_n - \theta)^2 \big) = b^2$ if $\theta = 0$.
Thus, at first sight, it may appear that $T_n$ performs better (in terms of mean squared estimation error) than $\overline{X}_n$
when the sample size $n$ is large. However, as Figure 2.1 on p.443 of Lehmann \& Casella 
(1998) shows, this apparent improvement in performance is misinformative: the supremum over $\theta$ of $n E \big( (T_n - \theta)^2 \big)$ approaches infinity as $n \rightarrow \infty$.
The problem with the analysis of $n E \big( (T_n - \theta)^2 \big)$ for each fixed $\theta$ 
as $n \rightarrow \infty$ is that this is a limit result that is pointwise in the 
parameter space $\Theta$. We should, instead, consider $n E \big( (T_n - \theta)^2 \big)$
across the entire parameter space $\Theta$ for each fixed $n$ and then let $n \rightarrow \infty$. 
As pointed out on p.153 of Hajek (1971):

\begin{quote}
	
Especially misinformative are those limit results that are not uniform. Then the limit
can exhibit some features that are not even approximately true for any finite $n$.
	
\end{quote}

\noindent and 

\begin{quote}
	
	Super efficient estimates produced by L.J. Hodges (see LeCam 1953, p.280)
	have their shocking properties only in the limit. For any finite $n$ they behave
	quite poorly for some parameter values. These values, however, depend on $n$ and 
	disappear in the limit.
	
\end{quote}

Kabaila (1995) presents the following confidence interval analogue of Hodges's superefficient estimator. Suppose that $X_1, X_2, \dots$ have the same probability distribution as before. Also define $\overline{X}_n$ and $T_n$ as before.
The usual $1-\alpha$ confidence interval for $\theta$ is 
$I_n = \big[\overline{X}_n - n^{-1/2} z_{1-\alpha}, 
\overline{X}_n + n^{-1/2} z_{1-\alpha} \big]$, where the quantile $z_a$ is defined by 
the requirement that $P(Z \le z_a) = a$ for $Z \sim N(0,1)$.
Of course,  $P_{\theta} (\theta \in I_n) = 1-\alpha$ for all $\theta$ and 
$n^{1/2} (\text{length of } I_n) = 2 z_{1-\alpha}$.
Let
\begin{equation*}
W_n =
\begin{cases}
1   &\text{if}\ \  |\overline{X}_n| > n^{-1/4}
\\
b^2 &\text{if}\ \ |\overline{X}_n| \le n^{-1/4},
\end{cases}
\end{equation*}
where, as before, $0 < b < 1$.
Now define the confidence interval
$J_n = \big[T_n - n^{-1/2} z_{1-\alpha} W_n^{1/2}, 
T_n + n^{-1/2} z_{1-\alpha}  W_n^{1/2} \big]$. 
It may be shown that for each $\theta$, 
$\lim_{n \rightarrow \infty} P_{\theta} (\theta \in J_n) = 1-\alpha$. In addition,
it may be shown that
$\lim_{n \rightarrow \infty} P_{\theta} (n^{1/2} (\text{length of } J_n) = 2 z_{1-\alpha} b) = 1$ for $\theta = 0$ and 
$\lim_{n \rightarrow \infty} P_{\theta} (n^{1/2} (\text{length of } J_n) = 2 z_{1-\alpha}) = 1$ for all $\theta \ne 0$.  Thus,
at first sight it may appear that the confidence interval $J_n$ performs better
than the confidence interval $I_n$ when $n$ is large.
Kabaila (1995) shows that this apparent improvement in performamce is misinformative:
the infimum over $\theta$ of $P_{\theta} (\theta \in J_n)$ approaches 0 as 
$n \rightarrow \infty$. In other words, the {\sl confidence coefficient} of $J_n$ approaches
0 as $n \rightarrow \infty$.
The problem with the analysis of $P_{\theta} (\theta \in J_n)$ for each fixed $\theta$ 
as $n \rightarrow \infty$ is that this is a limit result that is pointwise in the 
parameter space $\Theta$. We should, instead, consider $P_{\theta} (\theta \in J_n)$
across the entire parameter space $\Theta$ for each fixed $n$ and then let $n \rightarrow \infty$. 
This point is also made by Leeb \& P\"otscher (2005, pp.31--32).

We now present the main results of this section. Consider the linear regression model and 
parameter of interest $\theta = \boldsymbol{a}^{\top} \boldsymbol{\beta}$ described in the introduction. 
Remember, we assume that the last $p-q$ components of $\ba$ are zeros.
Also consider the 
MATA interval $I({\mathscr K}^*)$, with nominal coverage $1-\alpha$
and weight 
on model ${\cal M}_K$ proportional to $\exp(-\text{GIC}(K)/2)$,  
described in Section 4. 
The large sample framework that we consider is that
 $p$ and $q$ are fixed and $n \rightarrow \infty$. 
Of course, many of the quantities which were defined in Section 4
now depend on $n$. We make this dependence explicit in the notation
by using $\beta_{p,n}$, $v_{\theta, n}$, $v_{p, n}$, $\rho_n$, $d_n$ and $\gamma_n$
to denote  $\beta_p$, $v_{\theta}$, $v_{p}$, $\widetilde{\rho}$, $d$ and $\gamma$, 
respectively. 
Note that 
$v_{\theta,n}$, $v_{p,n}$ and $\rho_n$ are known, whereas $\gamma_n$ is an unknown parameter.
The main result of this section requires that the following assumption concerning $d_n$ holds.

\medskip

\noindent \underbar{Assumption A} \ Suppose that $\{ d_n \}$ is an increasing sequence of
nonnegative numbers that diverges to $\infty$ as $n \rightarrow \infty$. Also suppose that
$d_n / n \rightarrow 0$ as $n \rightarrow \infty$. 

\medskip

\noindent This assumption holds, for example, when $d_n = \ln(n)$, in which case the 
 weight on model ${\cal M}_K$ is proportional to
$\exp(-\text{BIC}(K)/2)$.

\bigskip

\noindent {\bf Theorem 3.} {\sl 
Consider the linear regression model and 
parameter of interest $\theta$ described in the introduction. Also consider the 
MATA interval $I({\mathscr K}^*)$, with nominal coverage $1-\alpha$
and weight 
on model ${\cal M}_K$ proportional to $\exp(-\text{GIC}(K)/2)$,  
described in Section 4. Here, 
${\mathscr K}^* = \big \{ \varnothing, \{ p \}   \big \}$.
Suppose that 
$p$ and $q$ are fixed
and 
that $\bD = \lim_{n \rightarrow \infty} \bX^{\top} \bX / n$ exists and is nonsingular.
Also suppose that \newline
$\ba^{\top} \bD^{-1} \bc \big / 
\big(\ba^{\top} \bD^{-1} \ba \, \bc^{\top} \bD^{-1} \bc \big)^{1/2} \ne 0$.
Finally, suppose that Assumption A holds. Then
\begin{enumerate}
	
	\item 
	
	The infimum over $\gamma_n \in \mathbb{R}$ of 	
	$P(\theta \in I({\mathscr K}^*))$ converges to 0,
	as $n \rightarrow \infty$.
	
	\item
	
	If $\beta_p$ and $\sigma^2$ ($\sigma^2 > 0$) are fixed and $\beta_p \ne 0$ then 
	$w(\varnothing;  {\mathscr K}^*)$ converges in probability to 1 and 
	$P(\theta \in I({\mathscr K}^*))$ converges to $1 - \alpha$,
	as $n \rightarrow \infty$.
	
	\item
	
	If $\beta_p$ and $\sigma^2$ ($\sigma^2 > 0$) are fixed and $\beta_p = 0$ then
		$w(\{p\};  {\mathscr K}^*)$ converges in probability to 1 and  
	$P(\theta \in I({\mathscr K}^*))$ converges to $1 - \alpha$,
	as $n \rightarrow \infty$.
	
\end{enumerate}
}

\medskip

\noindent This result is proved in the appendix. The most important part of this theorem is (a) which implies that the MATA interval, with  weight on model ${\cal M}_K$ proportional to
$\exp(-\text{BIC}(K)/2)$, has {\sl confidence coefficient} that approaches 0 as 
$n \rightarrow \infty$. In other words, this MATA interval should not be used when $n$ is large. Parts (b) and (c) of this theorem do not provide useful information as they are limits as $n \rightarrow \infty$ pointwise in the parameter space.

Another way of looking at Theorem 3 is the following. Consider the asymptotic framework that $\beta_p$ and $\sigma^2$ ($\sigma^2 > 0$) are both fixed.  If 
$\beta_p = 0$ then $\gamma_n =0$ and if $\beta_p > 0$ then $\gamma_n$ diverges to $\infty$ at rate $O(n^{1/2})$. Sequences $\gamma_n$ that diverge to $\infty$ at a slower rate are not included in this
analysis. The proof of part (a) of Theorem 3 presents one such sequence for which the coverage probability of the MATA interval $I({\mathscr K}^*)$ converges to 0. This sequence is ``missed'' in the asymptotic 
framework that $\beta_p$ and $\sigma^2$ are both fixed. 
In other words, this asymptotic framework does not lead to 
an accurate appreciation of the {\sl confidence coefficient} of this 
MATA interval  when $n$ is large.

We now turn our attention to the asymptotic framework that $m = n - p$ is fixed and
$n \rightarrow \infty$. The following result is proved in the appendix.

\bigskip

\noindent {\bf Theorem 4.} {\sl 
	Consider the linear regression model and 
	parameter of interest $\theta$ described in the introduction. Also consider the 
	MATA interval $I({\mathscr K}^*)$, with nominal coverage $1-\alpha$
	and weight 
	on model ${\cal M}_K$ proportional to $\exp(-\text{GIC}(K)/2)$,  
	described in Section 4. T
	Suppose that 
	$m = n - p$ is fixed.
	Also suppose that Assumption A holds. Then, for any given $\epsilon > 0$, 
	\begin{equation*}
	\sup_{\gamma} P\big(w_1(\widehat{\gamma}^2)\big) \ge \epsilon) \rightarrow 0 
	\quad \text{as} \quad n \rightarrow \infty.
	\end{equation*}
	In other words, $w_1(\widehat{\gamma}^2)$ converges in probability to 0
	as $n \rightarrow \infty$, uniformly in the parameter $\gamma$. 
}

\medskip

\noindent This theorem and its proof suggest that the MATA interval described in this
result will be close to the usual $1 - \alpha$ confidence interval for $\theta$ based
on the full model when $m = n - p$ is small compared to $n$. 
An interpretation of this suggested result is that this MATA interval is rather uninteresting when $m = n - p$ is small compared to $n$. 
A numerical exploration of the case that $m = n - p$ is small compared to $n$ is
presented in the Supplementary Material.

\bigskip

\noindent { \bf {7. CONCLUSION}}

\medskip

We have derived an easily-computed new upper bound on the  minimum coverage probability (i.e. the {\sl confidence coefficient}) of the MATA confidence  interval in the context of all possible subsets of a given set of explanatory variables in a linear regression model. The main application of this upper bound is that it can be used to help eliminate poorly performing model weights
from further consideration. In the Supplementary Material we present graphs similar to those displayed in Figure 1 for a wide range of values of $n$ and $p$. These graphs, 
combined with the large sample results presented in Section 6, show that the 
MATA confidence interval with weight 
on a model that is proportional to $\exp(-\text{BIC}/2)$, where
BIC is the Bayesian Information Criterion for this model, should not be used  
if $|\rho|_{\pkstyle{\rm max}}$ is not too far from 1 and $p/n$ is not too close to 1.

\bigskip

\noindent {\bf BIBLIOGRAPHY}

\medskip

\rf Abramowitz, M. \& Stegun, I.A. (1965). 
{\it Handbook of Mathematical Functions with 
	Formulas, Graphs, and Mathematical Tables}.
Dover, New York.

\rf Buckland, S.T., Burnham, K.P. \& Augustin, N.H. (1997). Model selection: an integral
part of inference. \textit{Biometrics}, 53, 603--618.

\rf Casella, G. \& Berger, R.L. (2002). {\it Statistical Inference, 2nd edition}.
Duxbury, Pacific Grove, CA.




\rf Chatterjee, S. \& Hadi, A.S. (2012). {\it Regression Analysis by Example, 5th edition}. 
Wiley, Hoboken, NJ.

\rf Fieberg, J. \& Johnson, D.H. (2015). MMI: Multimodel inference or models with management 
implications. {\it Journal of Wildlife Management}, 79, 708--718.

\rf Fletcher, D. \& Turek, D. (2011). Model-averaged profile likelihood confidence intervals.
{\it Journal of Agricultural,  Biological and  Environmental Statistics}, 17, 38--51.

\rf Graybill, F. A. (1976). {\it Theory and Application of the Linear Model}.
Duxbury, Pacific Grove CA.

\rf Hajek, J. (1971). Limiting properties of likelihoods and inference. In {\it Foundations of Statistical
	Inference: Proceedings of the Symposium on the Foundatuions of Statistical Inference prepared under the
	auspices of the Rene Descartes Foundation and held at the Department of Statistics, University of Waterloo,
	Ontario, Canada, from March 31 to April 9, 1970}, V.P. Godambe \& D.A. Sprott eds, pp. 142--159.
Holt, Reinhart and Winston, Toronto.

\rf Hjort, N.L. \& Claeskens, G. (2003). Frequentist model average estimators. 
{\it Journal of the American Statistical Association}, 98, 879--899.

\rf Johnson, N.L., Kotz, S. \& Balakrishnan, N. (1995). {\it Continuous Univariate Distributions, Volume 2, 2nd edition}. Wiley, New York.


\rf Kabaila, P. (1995). The effect of model selection on confidence regions and prediction regions.
{\it Econometric Theory}, 11, 537--549.

\rf Kabaila, P. \& Leeb, H. (2006). On the large-sample minimal coverage probability of confidence intervals after model
selection.
{\it Journal of the American Statistical Association}, 101, 619--629.

\rf Kabaila, P. \& Giri, K. (2009). Upper bounds on the minimum coverage probability of confidence intervals in regression after
model selection.
{\it Australian \& New Zealand Journal of Statistics},  51, 271--287.

\rf Kabaila, P., Welsh, A.H. \& Abeysekera, W. (2016). Model-averaged confidence intervals. 
{\it Scandinavian Journal of Statistics},
43, 35--48.

\rf Kabaila, P., Welsh, A.H. \& Mainzer, R. (2016). The performance of model averaged tail area confidence intervals. 
{\it Communications in Statistics - Theory and Methods}, 
DOI: 10.1080/03610926.2016.1242741

\rf LeCam, L. (1953). On some asymptotic properties of maximum likelihood estimates and related Bayes' estimates. University of California Press, p.277--328.

\rf Leeb, H. \& P\"otscher, B.M. (2005). Model selection and inference: facts and fiction.
{\it Econometric Theory}, 21, 21--59.

\rf Lehmann, E.L. \& Casella, G. (1998). {\it Theory of Point Estimation, 2nd edition}.
Springer, New York.

\rf Turek, D. \& Fletcher, D. (2012). Model-averaged Wald confidence intervals.
{\it Computational Statistics and Data Analysis},  56, 2809--2815.

\rf Wang, H. \& Zou, S.Z.F. (2013). Interval estimation by frequentist model averaging.
\textit{Communications in Statistics - Theory and Methods}, 42, 4342--4356.


\bigskip


\noindent  {\bf {APPENDIX}

\medskip

\noindent {\bf {The function $\boldsymbol{r}$ for weight on model $\boldsymbol{{\cal M}_K}$ proportional
to $\boldsymbol{\exp(-\text{GIC}(K)/2)}$}}}

\medskip

\noindent Suppose that 
\begin{equation*}
 w(K;  {\mathscr K}) 
= \frac{\exp(- \text{GIC}(K)/2)}{\sum_{L \in {\mathscr K}}\exp(- \text{GIC}(L)/2)},
\end{equation*}
where $\text{GIC}(K)$ is given by \eqref{GIC} for each $K \in {\mathscr K}$.
As noted in Appendix B of Kabaila \& Giri (2009), for each $K \in {\mathscr K}$,
\begin{equation}
\label{RSSKintermsofRSS}
\text{RSS}_K = \text{RSS} + U_K,  
\end{equation}
with the convention that $U_K = 0$ for 
$K = \varnothing$.
It follows from this that 
\begin{equation*}
 w(\varnothing;  {\mathscr K})
= \frac{1}
{1 + \displaystyle{\sum_{L \in {\mathscr K} \setminus \{\varnothing \}}  \left( 1 + \frac{U_L}{\text{RSS}}\right)^{-n/2} \exp \left( \frac{d |L|}{2}\right)}}
\end{equation*}
and, for $K \in {\mathscr K} \setminus \{\varnothing \}$,
\begin{equation*}
 w(K;  {\mathscr K}) 
= \frac{\displaystyle{ \left( 1 + \frac{U_K}{\text{RSS}}\right)^{-n/2} \exp \left( \frac{d |K|}{2}\right)}}
{1 +\displaystyle{ \sum_{L \in {\mathscr K} \setminus \{\varnothing \}} \left( 1 + \frac{U_L}{\text{RSS}}\right)^{-n/2} \exp \left( \frac{d |L|}{2}\right)}}.
\end{equation*}
It follows that $ w(K;  {\mathscr K}) $ is of the form \eqref{weight} for $r(x, y) = \exp(d \, y /2) \big/(1 + x)^{n/2}$,
where $r: (0, \infty) \times \{1, \dots, p-q\} \rightarrow (0, \infty)$ satisfies conditions \textbf{C1} and \textbf{C2}.


\medskip

\noindent { \bf {Proof of Lemma 1}}

\medskip

\noindent Suppose that $K$ is given ($K \in {\mathscr K}$). Let $\bbeta_K$ denote the $p$-vector obtained from
$\bbeta$ by setting to zero all of the components of $\bbeta$ with indices belonging to $K$. Since
$K$ is a subset of $\{q+1, \dots, p\}$, the first $q$ components 
of $\bbeta_K$
are $(\beta_1, \dots, \beta_q)$. 
Since we assume that the last $p - q$ components of $\ba$ are zeros, $\ba^{\top} \bbeta = \ba^{\top} \bbeta_K$. 
Thus
\begin{equation}
\label{ConsequenceLastComponentsZeros}
\frac{\boldsymbol{a}^{\top} \widehat{\boldsymbol{\beta}}_K - \ba^{\top} \bbeta}{S_K \, (v(K))^{1/2}} 
=\frac{\boldsymbol{a}^{\top} (\widehat{\boldsymbol{\beta}}_K - \bbeta_K)/\sigma}{(S_K/\sigma) \, (v(K))^{1/2}}.
\end{equation}
Since $\bH_K \bbeta_K = \bzero$,
\begin{equation*}
\bbeta_K
= \Big( \bI - (\bX^{\top} \bX)^{-1}   \bH_K^{\top} \big (\bH_K (\bX^{\top} \bX \big)^{-1} \bH_K^{\top})^{-1} \bH_K \Big)
\bbeta_K.
\end{equation*}
It follows from this and \eqref{FormulaBetaHatK} that
\begin{equation*}
\ba^{\top} \big(\widehat{\boldsymbol{\beta}}_K -  \bbeta_K\big) \big / \sigma
= \ba^{\top} \Big( \bI - (\bX^{\top} \bX)^{-1}   \bH_K^{\top}  \big(\bH_K (\bX^{\top} \bX \big)^{-1} \bH_K^{\top})^{-1} \bH_K \Big)
\big(\widehat{\bbeta} -  \bbeta_K\big) \big/ \sigma.
\end{equation*}
Obviously, 
$\big(\widehat{\bbeta} -  \bbeta_K \big ) \big/ \sigma
= \big(\widehat{\bbeta} -  \bbeta\big) \big/ \sigma
+ \big(\bbeta -  \bbeta_K\big) \big/ \sigma$.
Hence $\ba^{\top} \big(\widehat{\boldsymbol{\beta}}_K -  \bbeta_K\big) \big / \sigma$
can be expressed as a function of $\big(\widehat{\bbeta} -  \bbeta\big) \big/ \sigma$
and the variables 
in the set $\{ \beta_i / \sigma: i \in K \}$.

Now we turn our attention to the denominator of the right-hand side of \eqref{ConsequenceLastComponentsZeros}.
It follows from \eqref{RSSKintermsofRSS} that, for each $K \in {\mathscr K}$,
\begin{equation*}
\frac{\text{RSS}_K}{\sigma^2}=\frac{\text{RSS}}{\sigma^2}+V_K,
\end{equation*}
with the convention that 
$V_K = 0$ for 
$K = \varnothing$.
Hence, for each $K \in {\mathscr K}$,
\begin{equation*}
S_K / \sigma = \left(\frac{1}{n - p + |K|} \left(\frac{\text{RSS}}{\sigma^2}+V_K \right) \right)^{1/2}.
\end{equation*}
Suppose that $K \ne \varnothing$.
Note that $V_K$ can be expressed as a function of
 the random variables in the set
$\{ \widehat{\beta}_i / \sigma: i \in K \}$.
Therefore, $S_K / \sigma$ 
can be expressed as a function of
 ${\rm RSS} / \sigma^2$ 
and the random variables in the set
$\{ \widehat{\beta}_i / \sigma: i \in K \}$.
Hence \eqref{ConsequenceLastComponentsZeros}
can be expressed as a function of $(\widehat{\bbeta} - \bbeta) / \sigma$, ${\rm RSS} / \sigma^2$ and the 
random variables in the set
$\{ \widehat{\beta}_i / \sigma: i \in K \}$.
Since 
$\widehat{\beta}_i / \sigma = (\widehat{\beta}_i - \beta_i) / \sigma + \beta_i / \sigma$
for all $i \in K$, 
\eqref{ConsequenceLastComponentsZeros} can be 
expressed
as a function of $(\widehat{\bbeta} - \bbeta) / \sigma$, ${\rm RSS} / \sigma^2$ and the variables 
in the set $\{ \beta_i / \sigma: i \in K \}$. Also, for $K = \varnothing$,
\eqref{ConsequenceLastComponentsZeros}  can be 
expressed
as a function of $(\widehat{\bbeta} - \bbeta) / \sigma$ and ${\rm RSS} / \sigma^2$.

\bigskip


\noindent { \bf {Proof of Theorem 1(a)}}

\medskip

 \noindent It may be shown that, for given $\by$, $h \big(z, \by; {\mathscr K} \big) $ is a continuous decreasing function
of $z$. It follows from this that, for any given $z$,
\begin{equation*}
\Big \{ \widehat{\theta}_{\ell} \le z \le  \widehat{\theta}_u  \Big \} 
= \big \{ \alpha/2 \le h \big(z, \by; {\mathscr K} \big)  \le 1 - \alpha/2  \big \}.
\end{equation*}
Thus the coverage probability of the MATA 
interval
$I({\mathscr K})$, with nominal coverage $1-\alpha$, is
\begin{align}
\label{CovProb}
\notag
&P \big ( \alpha/2 \le h\big(\ba^{\top} \bbeta, \by;  {\mathscr K} \big) \le 1 - \alpha/2  \big ) 
\\
&= P \left( \frac{\alpha}{2}   
\le \sum_{K \in {\mathscr K}}  w(K;  {\mathscr K})  \,
G_{n - p + |K|} \left( \frac{\boldsymbol{a}^{\top} \widehat{\boldsymbol{\beta}}_K - \ba^{\top} \bbeta}{S_K \, (v(K))^{1/2}}  \right)
\le 1 - \frac{\alpha}{2}     \right).
\end{align}
We see from \eqref{weight} that, for each $K \in {\mathscr K}$, $w(K;  {\mathscr K}) $ is a function of 
$\text{RSS}/\sigma^2$ and $(1/\sigma) (\widehat{\beta}_{q+1}, \dots, \widehat{\beta}_p)$.
It follows from Lemma 1 that the vector of random variables in the set
\begin{equation*}
\left \{
G_{n - p + |K|} \left( \frac{\boldsymbol{a}^{\top} \widehat{\boldsymbol{\beta}}_K - \ba^{\top} \bbeta}{S_K \, (v(K))^{1/2}}  \right): K \in {\mathscr K}
\right\}
\end{equation*}
can be expressed
as a function of $(\widehat{\bbeta} - \bbeta) / \sigma$, ${\rm RSS} / \sigma^2$ and 
$(1/\sigma) (\beta_{q+1}, \dots, \beta_p)$.
Therefore
\begin{equation*}
\sum_{K \in {\mathscr K}}  w(K;  {\mathscr K})  \,
G_{n - p + |K|} \left( \frac{\boldsymbol{a}^{\top} \widehat{\boldsymbol{\beta}}_K - \ba^{\top} \bbeta}{S_K \, (v(K))^{1/2}}  \right)
\end{equation*}
can be expressed
as a function of $(\widehat{\bbeta} - \bbeta) / \sigma$, ${\rm RSS} / \sigma^2$ and 
$(1/\sigma) (\beta_{q+1}, \dots, \beta_p)$.

Now $(\widehat{\bbeta} - \bbeta) / \sigma$ and ${\rm RSS} / \sigma^2$ are independent random variables
with $(\widehat{\bbeta} - \bbeta) / \sigma \sim N \big(\bzero, (\bX^{\top} \bX)^{-1} \big)$ 
and ${\rm RSS} / \sigma^2 \sim \chi^2_{n-p}$. Hence \eqref{CovProb} is a function of 
$(1/\sigma) (\beta_{q+1}, \dots, \beta_p)$.

\newpage


\noindent { \bf  Proof of Theorem 2}

\medskip

\noindent Suppose that 
$(1/\sigma) (\beta_{\ell+1},\ldots, \beta_p)$ is given.
Choose
$\beta_{q+1}/\sigma= \cdots = \beta_{\ell}/\sigma = t$.
We will consider
$t \rightarrow \infty$.
Define 
${\mathscr J}$ to be the family of sets that belong to ${\mathscr K}$
and include at least one element of the set $\{q+1, \ldots, \ell\}$.
Remember, 
${\mathscr K}^{**}$ denotes the family of all subsets of $\{ \ell+1, \dots, p \}$,
including the empty set. Thus 
${\mathscr K} = {\mathscr J} \cup {\mathscr K}^{**}$,
where ${\mathscr J}$ and ${\mathscr K}^{**}$ are disjoint sets.
Hence
\begin{align}
\label{hInTwoParts}
\begin{split}
h \big(\ba^{\top} \bbeta, \by; {\mathscr K} \big) 
&= \sum_{K \in {\mathscr J}} w(K;  {\mathscr K}) \,
G_{n - p + |K|} \left( \frac{\boldsymbol{a}^{\top} \widehat{\boldsymbol{\beta}}_K - \ba^{\top} \bbeta}{S_K \, (v(K))^{1/2}}  \right) 
\\
&\ \ \ + \sum_{K \in {\mathscr K}^{**}} w(K;  {\mathscr K}) \,
G_{n - p + |K|} \left( \frac{\boldsymbol{a}^{\top} \widehat{\boldsymbol{\beta}}_K - \ba^{\top} \bbeta}{S_K \, (v(K))^{1/2}}  \right).
\end{split}
\end{align}

Now consider $K$ to be a given element of ${\mathscr J}$. It can be proved that $\big(\bH_K (\bX^{\top} \bX)^{-1} \bH_K^{\top} \big)^{-1}$
is a symmetric positive definite matrix. The noncentrality parameter $\lambda$, given by 
\eqref{NoncentralityParameter}, is bounded below by 
\begin{equation*}
(1/2) \, \| \bH_K (\bbeta /\sigma) \|^2 \,
\left(\text{smallest eigenvalue of  } \big(\bH_K (\bX^{\top} \bX)^{-1} \bH_K^{\top} \big)^{-1}  \right),
\end{equation*}
where $\| \cdot \|$ denotes the Euclidean norm. Since $K \in {\mathscr J}$ and
$\beta_{q+1}/\sigma= \cdots = \beta_{\ell}/\sigma = t$, $ \| \bH_K (\bbeta /\sigma) \|^2 \ge t^2$
and so $\lambda \rightarrow \infty$ as $t \rightarrow \infty$. Thus
\begin{equation*}
\displaystyle{\frac{V_K}{\text{RSS}/\sigma^2}}
\buildrel p \over \longrightarrow \infty 
\ \  \text{as}\ \  
\beta_{q+1}/\sigma= \cdots = \beta_{\ell}/\sigma = t \rightarrow \infty. 
\end{equation*}
It follows from condition C1 on the function $r$ that
\begin{equation}
\label{rTendToZero}
r \left( \displaystyle{\frac{V_K}{\text{RSS}/\sigma^2}}, |K| \right )
\buildrel p \over \longrightarrow 0 
\ \  \text{as}\ \  
\beta_{q+1}/\sigma= \cdots = \beta_{\ell}/\sigma = t \rightarrow \infty. 
\end{equation}

For each $K \in {\mathscr J}$,
\begin{equation}
\label{UpperBoundOn_w_K}
w(K; {\mathscr K})
= \displaystyle{\frac{r \left(\displaystyle{\frac{V_K}{\text{RSS}/\sigma^2}}, |K| \right)}
{1 + \displaystyle{\sum_{L \in {\mathscr K} \setminus \{\varnothing \}}} r \left(\displaystyle{\frac{V_L}{\text{RSS}/\sigma^2}}, |L| \right)}}
\le r \left(\displaystyle{\frac{V_K}{\text{RSS}/\sigma^2}}, |K| \right).
\end{equation}
Therefore, for
each $K \in {\mathscr J}$, $w(K; {\mathscr K}) \buildrel p \over \longrightarrow 0$, as 
$\beta_{q+1}/\sigma= \cdots = \beta_{\ell}/\sigma = t \rightarrow \infty$.
Since
\begin{equation*}
0 \le
\sum_{K \in {\mathscr J}} w(K; {\mathscr K}) \,
G_{n - p + |K|} \left( \frac{\boldsymbol{a}^{\top} \widehat{\boldsymbol{\beta}}_K - \ba^{\top} \bbeta}{S_K \, (v(K))^{1/2}}  \right)
\le \sum_{K \in {\mathscr J}} w(K; {\mathscr K}),
\end{equation*}
the first term on the right-hand side of \eqref{hInTwoParts} converges in probability to zero as 
$\beta_{q+1}/\sigma= \cdots = \beta_{\ell}/\sigma = t \rightarrow \infty$.

For $K \in {\mathscr K}^{**}$,
\begin{equation*}
\begin{split}
w(K;  {\mathscr K}) =
\begin{cases}
\displaystyle{\frac{1}
{\displaystyle{1 + \sum_{L \in {\mathscr J}} r\left (\frac{V_L}{\text{RSS}/\sigma^2}, |L| \right)
+ \sum_{L \in {\mathscr K}^{**} \setminus \{\varnothing \}} r\left (\frac{V_L}{\text{RSS}/\sigma^2}, |L| \right)}} }
&\text{for}\ \ \ K = \varnothing \\
\\
\displaystyle{\frac{r\left (\displaystyle{\frac{V_K}{\text{RSS}/\sigma^2}}, |K| \right)}
{\displaystyle{1 + \sum_{L \in {\mathscr J}} r\left (\frac{V_L}{\text{RSS}/\sigma^2}, |L| \right)
+ \sum_{L \in {\mathscr K}^{**} \setminus \{\varnothing \}} r\left (\frac{V_L}{\text{RSS}/\sigma^2}, |L| \right)}} } 
&\text{otherwise}.
\end{cases}
\end{split}
\end{equation*}
It follows from \eqref{rTendToZero} that, for each $K \in {\mathscr K}^{**}$,
\begin{equation*}
w(K;  {\mathscr K}) 
\buildrel p \over \longrightarrow w(K;  {\mathscr K}^{**}) 
\ \  \text{as}\ \  
\beta_{q+1}/\sigma= \cdots = \beta_{\ell}/\sigma = t \rightarrow \infty. 
\end{equation*}
It follows from \eqref{hInTwoParts} that
\begin{equation}
\label{ApproxTo_h}
h \big(\ba^{\top} \bbeta, \by; {\mathscr K} \big) 
- h \big(\ba^{\top} \bbeta, \by; {\mathscr K}^{**} \big) 
\buildrel p \over \longrightarrow 0
\ \  \text{as}\ \  
\beta_{q+1}/\sigma= \cdots = \beta_{\ell}/\sigma = t \rightarrow \infty. 
\end{equation}

By Theorem 1,  the coverage probability of 
the MATA 
interval $I({\mathscr K})$, with nominal coverage $1-\alpha$, 
is a function of 
$(1/\sigma)(\beta_{q+1},\ldots, \beta_{\ell})$
and $(1/\sigma)(\beta_{\ell+1},\ldots, \beta_p)$.
Since we suppose that 
$(1/\sigma) (\beta_{\ell+1},\ldots, \beta_p)$ is given,
the infimum of this coverage probability over
$(1/\sigma)(\beta_{q+1},\ldots, \beta_{\ell})$
and $(1/\sigma)(\beta_{\ell+1},\ldots, \beta_p)$
is less than or equal to
\begin{equation*}
P \big ( \alpha/2 \le h\big(\ba^{\top} \bbeta, \by;  {\mathscr K} \big) \le 1 - \alpha/2  \big ) 
\end{equation*}
for every $(1/\sigma)(\beta_{q+1},\ldots, \beta_{\ell}) \in {\mathbb R}^{\ell - q}$. Also, it follows from
\eqref{ApproxTo_h} that
\begin{equation*}
P \big ( \theta \in I({\mathscr K}) )
=P \big ( \alpha/2 \le h\big(\ba^{\top} \bbeta, \by;  {\mathscr K} \big) \le 1 - \alpha/2  \big ) 
\end{equation*}
approaches 
\begin{equation*}
P \big ( \alpha/2 \le h\big(\ba^{\top} \bbeta, \by;  {\mathscr K}^{**} \big) \le 1 - \alpha/2  \big ) 
= P \big ( \theta \in I({\mathscr K}^{**}) )
\end{equation*}
as $\beta_{q+1}/\sigma= \cdots = \beta_{\ell}/\sigma = t \rightarrow \infty$.
Therefore the infimum of the coverage probability of 
the MATA 
interval $I({\mathscr K})$, with nominal coverage $1-\alpha$, is less than or equal to
\begin{equation}
\label{ApproxToCP}
P \big ( \theta \in I({\mathscr K}^{**}) ).
\end{equation}
Since this is true for every given 
$(1/\sigma)(\beta_{q+1},\ldots, \beta_{\ell}) \in {\mathbb R}^{\ell - q}$,
the infimum of the coverage probability of 
the MATA 
interval $I({\mathscr K})$, with nominal coverage $1-\alpha$, is less than or equal to
the minimum over $(1/\sigma)(\beta_{q+1},\ldots, \beta_{\ell}) \in {\mathbb R}^{\ell - q}$
of \eqref{ApproxToCP}.

\newpage

\noindent { \bf  Proof of Theorem 3}

\medskip

\noindent Consider the MATA interval described in the statement of the theorem and suppose that the assumptions made in this statement hold. It follows that the sequence
$\{ \rho_n \}$ converges to the non-zero number 
$\rho_{\infty} = \ba^{\top} \bD^{-1} \bc \big / 
\big(\ba^{\top} \bD^{-1} \ba \, \bc^{\top} \bD^{-1} \bc \big)^{1/2}$
as $n \rightarrow \infty$. 
Let 
$\widehat{\gamma}_n = \widehat{\beta}_p / \big(\widehat{\sigma} \, v_{p,n}^{1/2} \big)$. 
As in Section 4, define the function $w_1$ by \eqref{Definition_w1}.
It follows from p.40 of Kabaila, Welsh \& Abeysekera (2016) that the function defined by 
\eqref{hzyK} is given by 
\begin{align*}
h \big(z, \by; {\mathscr K}^* \big) 
&= w(\{p\};  {\mathscr K}^*) \,
G_{m+1} \left( \left(\frac{m+1}{m +\widehat{\gamma}_n^2}\right)^{1/2} \, 
\frac{\widehat{\theta} - v_{\theta,n}^{1/2} \, \widehat{\sigma} \, \rho_n \, \widehat{\gamma}_n - z}
{v_{\theta,n}^{1/2} \, \widehat{\sigma} \, (1 - \rho_n^2)^{1/2}} \right)
\\
&\ \ \ \ \ + w(\varnothing;  {\mathscr K}^*) \, G_m \left(\frac{\widehat{\theta} - z}{\widehat{\sigma} \, v_{\theta,n}^{1/2}}  \right),
\end{align*}
where $w(\{p\};  {\mathscr K}^*) = w_1(\widehat{\gamma}_n^2)$
and $w(\varnothing;  {\mathscr K}^*) = 1 - w_1(\widehat{\gamma}_n^2)$.
Remember, the MATA interval is obtained by solving the equations \eqref{MATA_EstimatingEqns}. Since, for any given $\by$, $h \big(z, \by; {\mathscr K}^* \big)$ 
is a continuous
decreasing function of $z \in \mathbb{R}$, the coverage probability of the MATA 
interval $I({\mathscr K}^*)$ is
\begin{equation}
\label{CP_MATA_interval}
P(\theta \in I({\mathscr K}^*)) 
=  \big( \alpha/2 \le h \big(\theta, \by; {\mathscr K}^* \big) \le 1 - \alpha/2 \big).
\end{equation}
We will need the following consequence of the exponential inequality
4.4.26 on p.70 of Abramowitz \& Stegun (1965):
\begin{equation}
\label{ExponentialIneq}
\frac{1}{1 + \exp \displaystyle{\left( \frac{z}{2} . \,  \frac{n}{m} - \frac{d_n}{2} \right)}}
< w_1(z) <
\frac{1}{1 + \exp \displaystyle{\left( \frac{z}{2} . \,  \frac{n}{z + m} - \frac{d_n}{2} \right)}},
\end{equation}
for all $z > 0$. 


\medskip

\noindent \underbar{Proof of part (a)}

\medskip

\noindent We show that the coverage probability of the interval $I({\mathscr K}^*)$ converges to 0 when we consider $\sigma^2 > 0$ to be fixed and 
that 
$\beta_{p,n} = \sigma \, (v_{p,n} \, d_n / 2)^{1/2}$. It follows from this 
that $\gamma_n^2 = d_n / 2$. 
Now
\begin{equation*}
\widehat{\gamma}_n^2 = \frac{1}{\widehat{\sigma}^2 / \sigma^2} \, B_n^2,
\end{equation*}
where $B_n = \widehat{\beta}_p / (\sigma \, v_{p,n}^{1/2})$.
Note that $B_n \sim N \big((d_n/2)^{1/2}, 1 \big)$.
It follows from this and Assumption A that 
$\widehat{\gamma}_n^2 = (d_n/2) + O_p \big(d_n^{1/2}\big)$.
Hence, by the first inequality in \eqref{ExponentialIneq}, $w_1(\widehat{\gamma}_n^2) \buildrel p \over \longrightarrow 1$,
where $\buildrel p \over \longrightarrow$ denotes convergence in probability as 
$n \rightarrow \infty$. 
It follows from the fact that $0 < G_m(z) < 1$ for all 
$z \in \mathbb{R}$ that 
\begin{equation*}
(1 - w_1(\widehat{\gamma}_n^2)) \, G_m \left(\frac{\widehat{\theta} - \theta}{\widehat{\sigma} \, v_{\theta,n}^{1/2}}  \right)
\buildrel p \over \longrightarrow 0.
\end{equation*}
Now
\begin{equation}
\label{ArgumentGmplus1}
\left(\frac{m+1}{m +\widehat{\gamma}_n^2}\right)^{1/2} \, 
\frac{\widehat{\theta} - v_{\theta,n}^{1/2} \, \widehat{\sigma} \, \rho_n \, \widehat{\gamma}_n - \theta}
{v_{\theta,n}^{1/2} \, \widehat{\sigma} \, (1 - \rho_n^2)^{1/2}}
= \left(\frac{m+1}{m +\widehat{\gamma}_n^2}\right)^{1/2} \, \frac{1}{\widehat{\sigma} / \sigma} \, \frac{A_n - \rho_n \, B_n}{(1 - \rho_n^2)^{1/2}},
\end{equation}
where 
$A_n = \big(\widehat{\theta} - \theta \big) \big/ \big(\sigma \, v_{\theta,n}^{1/2}\big)$.
By Assumption A and since $\widehat{\gamma}_n^2 = (d_n/2) + O_p \big(d_n^{1/2}\big)$,
\begin{equation*}
\left(\frac{m+1}{m +\widehat{\gamma}_n^2}\right)^{1/2}
\buildrel p \over \longrightarrow 1.
\end{equation*}
Obviously, $\widehat{\sigma} / \sigma \buildrel p \over \longrightarrow 1$.
Since
\begin{align*}
\label{JointPdfAnBn}
\left[ {\begin{array}{c}
	A_n \\
	B_n
	\end{array} } \right] \sim N \left(
\left[ {\begin{array}{c}
	0 \\
	\gamma_n
	\end{array} } \right],
\left[ {\begin{array}{cc}
	1 & \rho_n  \\
	\rho_n  & 1
	\end{array} } \right]
\right),
\end{align*}
\begin{equation*}
\frac{A_n - \rho_n \, B_n}{(1 - \rho_n^2)^{1/2}} \sim
N \left( - \frac{\rho_n}{(1 - \rho_n^2)^{1/2}}  \left(\frac{d_n}{2} \right)^{1/2}, \, 1 \right).
\end{equation*}
As noted earlier, $\rho_n$ converges to $\rho_{\infty} \ne 0$, as $n \rightarrow \infty$.
We have the following two cases to consider. If $\rho_{\infty} > 0$ then
$G_{m+1}$, evaluated at the right-hand side of \eqref{ArgumentGmplus1}, converges in probability to 0, as $n \rightarrow \infty$.  If, on the other hand, $\rho_{\infty} < 0$ then
$G_{m+1}$, evaluated at the right-hand side of \eqref{ArgumentGmplus1}, converges in probability to 1, as $n \rightarrow \infty$. 
Consequently, if $\rho_{\infty} > 0$ then 
$h \big(\theta, \by; {\mathscr K}^* \big) \buildrel p \over \longrightarrow 0$
and if $\rho_{\infty} < 0$ then 
$h \big(\theta, \by; {\mathscr K}^* \big) \buildrel p \over \longrightarrow 1$.
It follows from \eqref{CP_MATA_interval} that 
	$P(\theta \in I({\mathscr K}^*))$ converges to 0,
as $n \rightarrow \infty$,
in both of these cases. 

\medskip

\noindent \underbar{Proof of part (b)}

\medskip

\noindent Suppose that $\beta_p$ and $\sigma^2$ ($\sigma^2 > 0$) are fixed and $\beta_p \ne 0$. Now
\begin{equation*}
\gamma_n = n^{1/2} \frac{\beta_p}{\sigma (n \, v_{p,n})^{1/2}}
\end{equation*}
and 
$n \, v_{p,n} = n \, \bc^{\top} (\bX^{\top} \bX)^{-1} \bc 
= \bc^{\top} (\bX^{\top} \bX / n)^{-1} \bc \rightarrow \bc^{\top} \bD^{-1} \bc$,
as $n \rightarrow \infty$.
Thus $\gamma_n^2 = O(n)$. Now 
\begin{equation*}
\widehat{\gamma}_n^2 = \frac{1}{\widehat{\sigma}^2 / \sigma^2} \, B_n^2,
\end{equation*}
where $B_n = \widehat{\beta}_p / (\sigma \, v_{p,n}^{1/2}) \sim N \big(\gamma_n, 1 \big)$.
By the second inequality in 
\eqref{ExponentialIneq}, $w_1(\widehat{\gamma}_n^2) \buildrel p \over \longrightarrow 0$.
Thus $w(\varnothing;  {\mathscr K}^*) = 1 - w_1(\widehat{\gamma}_n^2) \buildrel p \over \longrightarrow 1$.

It follows from the fact that 
$0 < G_{m+1}(z) < 1$ for all 
$z \in \mathbb{R}$ that 
\begin{equation*}
w_1(\widehat{\gamma}_n^2) \, G_{m+1} \left(\left(\frac{m+1}{m +\widehat{\gamma}_n^2}\right)^{1/2} \, 
\frac{\widehat{\theta} - v_{\theta,n}^{1/2} \, \widehat{\sigma} \, \rho_n \, \widehat{\gamma}_n - \theta}
{v_{\theta,n}^{1/2} \, \widehat{\sigma} \, (1 - \rho_n^2)^{1/2}} \right)
\buildrel p \over \longrightarrow 0.
\end{equation*}
Since 
\begin{equation*}
\frac{\widehat{\theta} - \theta}{\widehat{\sigma} \, v_{\theta,n}^{1/2}} \sim t_m,
\end{equation*}
\begin{equation*}
G_m \left(\frac{\widehat{\theta} - \theta}{\widehat{\sigma} \, v_{\theta,n}^{1/2}} \right)
\sim U(0,1),
\end{equation*}
for each $m$,
where $U(0,1)$ denotes the uniform distribution on the interval $(0,1)$. By Slutsky's theorem, 
$h \big(\theta, \by; {\mathscr K}^* \big) \buildrel d \over \longrightarrow U(0,1)$,
where $\buildrel d \over \longrightarrow$ denotes convergence in distribution, as 
$n \rightarrow \infty$. It follows from \eqref{CP_MATA_interval} that 
$P(\theta \in I({\mathscr K}^*)) \rightarrow 1 - \alpha$,
as $n \rightarrow \infty$.

\medskip

\noindent \underbar{Proof of part (c)}

\medskip

\noindent Suppose that $\beta_p$ and $\sigma^2$ ($\sigma^2 > 0$) are fixed and 
$\beta_p = 0$. In this case $\widehat{\gamma}_n^2 = O_p(1)$ and, by the first inequality in 
\eqref{ExponentialIneq},
$w_1(\widehat{\gamma}_n^2) \buildrel p \over \longrightarrow 1$. Thus
\begin{equation*}
(1 - w_1(\widehat{\gamma}_n^2)) \, G_m \left(\frac{\widehat{\theta} - \theta}{\widehat{\sigma} \, v_{\theta,n}^{1/2}}  \right)
\buildrel p \over \longrightarrow 0.
\end{equation*}
Since
\begin{equation*}
\left(\frac{m+1}{m +\widehat{\gamma}_n^2}\right)^{1/2} \, 
\frac{\widehat{\theta} - v_{\theta,n}^{1/2} \, \widehat{\sigma} \, \rho_n \, \widehat{\gamma}_n - \theta}
{v_{\theta,n}^{1/2} \, \widehat{\sigma} \, (1 - \rho_n^2)^{1/2}}
\sim t_{m+1},
\end{equation*}
\begin{equation*}
G_{m+1} \left(\left(\frac{m+1}{m +\widehat{\gamma}_n^2}\right)^{1/2} \, 
\frac{\widehat{\theta} - v_{\theta,n}^{1/2} \, \widehat{\sigma} \, \rho_n \, \widehat{\gamma}_n - \theta}
{v_{\theta,n}^{1/2} \, \widehat{\sigma} \, (1 - \rho_n^2)^{1/2}} \right)
\sim U(0,1),
\end{equation*}
for each $m$. By Slutsky's theorem, 
$h \big(\theta, \by; {\mathscr K}^* \big) \buildrel d \over \longrightarrow U(0,1)$. It follows from \eqref{CP_MATA_interval} that 
$P(\theta \in I({\mathscr K}^*)) \rightarrow 1 - \alpha$,
as $n \rightarrow \infty$.

\bigskip

\noindent { \bf  Proof of Theorem 4}

\medskip

\noindent Obviously, $w_1(\widehat{\gamma}_n^2)$ is a decreasing function of
$\widehat{\gamma}_n^2$. Now $\widehat{\gamma}_n^2$
has the same distribution as
\begin{equation*}
\frac{U}{Q/m},
\end{equation*}
where $U$ and $Q$ are independent, $U$ has a noncentral $\chi^2$ distribution with 1
degree of freedom and noncentrality parameter $\gamma^2$ and $Q$ has a 
$\chi_m^2$ distribution. For every $c > 0$, 
\begin{equation*}
P \left(\frac{U}{Q/m} \le c \right)
\end{equation*}
is a decreasing function of $\gamma^2$, see e.g. Johnson, Kotz \& Balakrishnan (1995, p.487).
Suppose that $\epsilon > 0$ is given. This result implies that  
\begin{equation*}
\sup_{\gamma} P_{\gamma} \big(w_1(\widehat{\gamma}_n^2) \ge \epsilon \big)
= P_{\gamma = 0} \big(w_1(\widehat{\gamma}_n^2) \ge \epsilon \big),
\end{equation*}
where $P_{\gamma}$ denotes the probability for true parameter value $\gamma$. Obviously,
\begin{equation*}
w_1(\widehat{\gamma}_n^2)
= \frac{1}{1 + \exp \left( \displaystyle{ \frac{n}{2} \left(\ln \left(1 + \frac{\widehat{\gamma}_n^2}{m} \right) - \frac{d_n}{n}  \right) } \right)}.
\end{equation*}
Suppose that $\gamma = 0$, so that $\widehat{\gamma}_n^2$
has a $\chi_1^2$ distribution. By Assumption A, 
$d_n / n \rightarrow 0$ as $n \rightarrow \infty$. Therefore
$P_{\gamma = 0} \big(w_1(\widehat{\gamma}_n^2) \ge \epsilon \big) \rightarrow 0$
as $n \rightarrow \infty$.

\end{document}